# Graphene Metallization of High-Stress Silicon Nitride Resonators for Electrical Integration

Sunwoo Lee[1, a)], Vivekananda P. Adiga[2, a)], Robert A. Barton[3, a)], Arend van der Zande[3], Gwan-Hyoung Lee[4,5], B. Rob Ilic[6], Alexander Gondarenko[4], Jeevak M. Parpia[2], Harold G. Craighead[2], James Hone[4, b)]

[1]*Department of Electrical Engineering, Columbia University, New York, New York 10027, USA.*
[2]*School of Applied and Engineering Physics, Cornell University, Ithaca, New York 14853, USA.*
[3]*Energy Frontier Research Center, Columbia University, New York, New York 10027, USA.*
[4]*Department of Mechanical Engineering, Columbia University, New York, New York 10027, USA.*
[5]*Samsung-SKKU Graphene Center (SSGC), Suwon, Gyeonggi 440-746, Korea.*
[6]*The Cornell NanoScale Science & Technology Facility (CNF), Ithaca, New York 14853, Korea.*

High stress stoichiometric silicon nitride resonators, whose quality factors exceed one million, have shown promise for applications in sensing and signal processing. Yet, electrical integration of the insulating silicon nitride resonators has been challenging, as depositing even a thin layer of metal degrades the quality factor significantly. In this work, we show that graphene used as a conductive coating for $Si_3N_4$ membranes reduces the quality factor by less than 30 % on average, which is minimal when compared to the effect of conventional metallization layers such as chromium or aluminum. The electrical integration of $Si_3N_4$-Graphene (SiNG) heterostructure resonators is demonstrated with electrical readout and electro-static tuning of the frequency by up to 1 % per volt. These studies demonstrate the feasibility of hybrid graphene/nitride mechanical resonators in which the electrical properties of graphene are combined with the superior mechanical performance of silicon nitride.

Resonators are essential components in modern electronics, and are often used as filters or oscillators. A key feature of a good resonator is a high quality factor, which conventional electronic filters have failed to meet. Nanoelectromechanical systems (NEMS) are of interest for sensing and signal processing applications because they can have the required high Q in addition to tunability and the potential for integration with silicon electronics[1,2,3,4]. Recently, stoichiometric silicon nitride resonators have been studied for their extremely high quality factor that can exceed one million, which originates from the high stress they possess[5,6,7]. However, the insulating nature of the material has hindered its broader implementation. Unfortunately, deposition of conventional metals as conducting layers degrades quality factor[8,9,10] – by more than a factor of four for only 5nm of chromium[8], and even more severely for thicker layers[11]. Therefore, there is a strong motivation to identify conducting materials that do not impact the quality factor. In order for the dissipation to be dominated by the silicon nitride itself, the

---

metallic coating needs to be much thinner than $Si_3N_4$ film, which is less than 20nm for the highest quality factor demonstrated to date[6].

In this work, we show that graphene can be used as an ideal conductive coating for $Si_3N_4$ membranes. Chemical vapor deposition (CVD) graphene[12] is directly transferred on top of the suspended $Si_3N_4$ resonators using a dry polydimethyl siloxane (PDMS) – poly(methyl methacrylate) (PMMA) transfer technique[13]. We have found that the addition of graphene layer only reduces the quality factor of the $Si_3N_4$ resonators by less than 30 %, in contrast with more than a factor of 4 reported previously using chromium[8]. Furthermore, the graphene coating allows electrical actuation and detection of $Si_3N_4$ resonators. Both the capacitive readout scheme as well as the direct detection scheme based on graphene's transconductance[14] are used to demonstrate electrical integration. In addition, electrostatic tuning of the resonant frequency close to 10 % is observed, at the cost of increased dissipation at high gate bias due to displacement current as reported previously[15], suggesting the trade-off between the frequency tuning and the quality factor degradation. These studies pave the way for electrically integrated, mechanically resonant filters and oscillators with quality factor approaching one million at room temperature.

Figure 1 shows a schematic of the fabrication process. We use regular silicon substrates (5 – 10 Ω-cm). After standard RCA clean[16] on the substrate, ~ 660 nm of $SiO_2$ is thermally grown. On top of the oxide, ~ 110 nm of high-stress stoichiometric $Si_3N_4$ is grown via low-pressure chemical vapor deposition (LPCVD). Following the deposition, PMMA is spun on the substrate to pattern a circular array of holes with diameter close to 500 nm using electron beam lithography (EBL, JEOL 6300). The PMMA pattern is then transferred onto the underlying nitride using $CF_4$ plasma etching (OXFORD 80) as shown in Fig. 1(b). The etched structure is then immersed in 49% hydrofluoric acid (HF) for 75 seconds to etch away the underlying $SiO_2$ to suspend the $Si_3N_4$ drum as shown in Fig. 1(c). After the HF etching, the sample is dried using critical point dryer (CPD, Tousimis) in order to avoid stiction due to capillary forces from phase transition[17].

To further demonstrate the applicability of this technique for wafer scale integration we use CVD graphene with vastly different grain sizes[12,18], whose electrical and mechanical properties can be close to those of mechanically exfoliated graphene[19,20]. Detailed growth conditions can be found at supplementary information. Following the graphene growth, a layer of poly(methyl methacrylate) (PMMA) is spin-coated on top of the graphene/copper stack, then a PDMS stamp is pressed on top to support the PMMA/graphene stack during etching and transfer[13]. After wet-etch of the copper with ammonium persulfate (Transene, APS100, 20 Wt%), the graphene/PMMA/PDMS stack is gently pressed against the aforementioned substrate with suspended $Si_3N_4$ drums, which is plasma treated to promote adhesion between the graphene and the nitride (Plasma Etch PE-50). Upon heating the stamped substrate at 170 °C for one minute, the PDMS is slowly peeled off. Finally, in order to remove the capping PMMA without collapsing the suspended drum, the sample is annealed at 345 °C for 6 hours in forming gas (5 % hydrogen and 95 % argon by volume)[21]. Figure 1 (e) and (f) respectively show an optical image and a false-color SEM image of the fabricated SiNG. The colored areas of (f) denote the circular array of the nano-holes that define the size of the suspended SiNG drum resonators, and it was found by SEM analysis that the graphene does not cover the nano-holes completely due to the occasional rips over the holes.

In this study, we used two different schemes to measure mechanical resonance. First, as shown in Fig 2(a), is a piezo-drive/optical detection scheme. A 633 nm HeNe laser is used to probe the motion of the resonators; variations in the reflected light intensity are used to transduce mechanical motion into an optical signal, where the mechanical drive is accomplished using a piezoelectric element. The second method is an electrical-drive/electrical detection scheme as shown in Fig. 2(b). In this method, the resonant motion is actuated by the electrical signal coming from the silicon back-gate, while the induced vibration is translated into the change in gate-to-membrane capacitance, generating radio frequency (RF) signal out to source. All the measurements are done under vacuum (~$10^{-6}$ Torr) to minimize air damping, and the devices were kept under vacuum for several hours prior to measurements to remove moisture and chemical residues[14].

In order to examine the change in quality factor for $Si_3N_4$ drums with and without graphene on top, we have measured the first few modes of $Si_3N_4$ drums before and after the graphene transfer. Figure. 3(a) shows the change in quality factor before and after the graphene transfer for the first 5 modes of a single 206 μm diameter drum. While we observe the quality factor enhancement at higher order modes due to the destructive interference of elastic waves as previously reported[22,23], the quality factor degradation is smaller compared to previous metallization schemes[8,9]. To establish more robust statistics, Figure. 3(b) demonstrates the quality factor degradation in the first five modes of 12 different devices. The average quality factor degradation due to the graphene metallization of $Si_3N_4$ resonator is less than 30 % as shown by the green line. In addition, we have found that the degradation does not depend on graphene quality in terms of grain size as both large grain and small grain graphene caused a similar amount of dissipation. On a few occasions, the quality factors have increased after the graphene deposition, which is possibly due to the removal of polymer impurity during forming gas annealing or opto-mechanical effect due to the bi-morphic nature of the SiNG resonator[24]. However, since the quality factor does not change for different detection laser power, the opto-mechanical justification is unlikely.

Our finding that graphene does not significantly degrade the quality factor of the underlying resonator is surprising, given that mechanical devices made from graphene alone have for the most part shown poor quality factors at room temperature[25,26]. By subtracting dissipation (1/Q) of bare $Si_3N_4$ resonator from the dissipation of SiNG, we can isolate the dissipation from the graphene and calculate the damping rate ($\gamma = f/Q$) for higher order modes of different device sizes. Figure 3(c) represents the graphene damping rate as a function of device dimension, and appears to show that the damping caused by graphene decreases monotonically with device size. This suggests that the dissipation in graphene mechanical systems can be studied by this technique. Further investigation is needed to model the cause of the dissipation due to graphene in this experiment, yet there are several possible explanations why graphene is superior to other metals as a coating such as graphene's extremely low thickness and large grain size compared to that of evaporated metals. However, the fact that we have not seen noticeable differences in graphene-induced dissipation for large and small grain graphene indicates graphene's low thickness might be the main attribute responsible for low added dissipation. This measurement also sheds light on previous studies of quality factors in bare graphene[27]. In comparison to resonators made solely from graphene, heterostructure devices such as these are useful as a means to raise the total energy stored in a device while maintaining the beneficial properties of graphene.

Having found that graphene does not significantly degrade the performance of the nitride resonators, electrical measurements are performed to further study the electrical characteristics of the SiNG heterostructure as shown in Fig. 2 (b). Source and drain electrodes are deposited using a shadow mask, and the actuation method is changed from pierzo-drive to electrical-drive. Vector network analyzer (VNA) output and DC biasing source (Yokogawa, GS200) are combined through a bias tee (Mini-circuits, ZFBT-4GW+) and connected to the silicon global back-gate through wire bonding. In order to examine the power handling of the SiNG resonator, we have swept input power until the resonance becomes nonlinear. Figure 4(a) shows that, as the input RF power increases above -40 dBm at $V_g$ = 0 V, the SiNG drum starts to resonate and, at 0 dBm, resonance becomes non-linear, exhibiting the bi-stability, defining the dynamic range of the resonator.

Another important characteristic of NEMS resonator is frequency. Figure. 4(b) shows how the resonant frequency of the fundamental mode in 124 μm SiNG drum changes as a function of gate bias. As expected from the high-stress nature of the silicon nitride film, the tunability is smaller compared to that of graphene resonators (on the order of a few hundred percent)[28], and spring constant softening due to nonlinear electrostatic interaction[29] is observed. As the gate voltage is swept up to 30 V, the resonant frequency of SiNG drum drops from 2.77 MHz down to 2.50 MHz, with tunability of about 0.3 % per volt. In addition to the frequency tuning from applied electrostatic force, Figure 4(b) also shows how the quality factor decreases as a function of gate bias. Over the change of 30 V in gate bias, the quality factor decreased by more than an order of magnitude, even though the input power was adjusted at each bias point to keep the resonant response Lorentzian. Such drop in the quality factor with increasing gate bias has been reported[15], but not in such extreme magnitude. The reason is that for SiNG drums studied here, the suspended device area is much larger (by more than a factor of hundred) compared to the reference, while the suspension height is rather comparable (larger only by about factor of two), making SiNG drums suffer from much larger displacement current for the same electrostatic force from the gate. With such differences taken into account, we were able to fit the data into the model given by the reference[15] which is shown in Fig. 4(b) as a red dashed line. More details for the model and the fit can be found in the supplementary information.

Furthermore, we have both actuated and detected the SiNG resonators electrically, by using capacitance as well as a direct detection scheme[14]. Figure 5(a) shows how the resonance of the 91 μm diameter SiNG drum changes with applied gate bias, where the capacitive detection scheme is used. On the other hand, Figure 5(b) shows the resonance as a function of the gate bias using the direct detection scheme with applied source-drain bias of 9 V. We have found that the capacitive signal is dominant in this system owing to the large surface area and large graphene-to-gate distance, and applied source-drain bias for direct detection increases the signal by only a few dB. Figure 5(b) also shows that the downshifted resonance and the increased tuning range due to the expansion from the joule heating on the membrane. With reduced graphene-to-gate distance, it is expected that the resonant signal will be even further amplified owing to the graphene's transconducting nature[30].

In summary, we demonstrated a novel fabrication process for metallizing high-stress stoichiometric $Si_3N_4$ resonators using graphene. We showed that the addition of a graphene layer onto $Si_3N_4$ drum resonator does not significantly degrade the quality factor of $Si_3N_4$, and we used the high-quality factor $Si_3N_4$ resonators to sensitively measure the mechanical dissipation in the graphene layer. In addition, the graphene layer allowed the resonance to be actuated electrically as well as tuned with applied electrostatic force. Lastly, displacement current induced quality factor degradation has been confirmed, suggesting design trade-offs between device size, quality factor, and frequency tuning range. With the new fabrication technique and electrical properties of SiNG drums in mind, further improvement in the direct electrical readout scheme should enable on-chip high quality factor resonators that are capable of replacing off-chip frequency references such as quartz crystals.

The authors like to thank Nicholas Petrone, Adam Hurst, Michael Lekas, Changyao Chen, and Victor Abramsky for critical discussions. Fabrication was performed at the Cornell Nano-Scale Facility, a member of the National Nanotechnology Infrastructure Network, which is supported by the National Science Foundation (Grant ECS-0335765), and Center for Engineering and Physical Science Research (CEPSR) Clean Room at Columba University. The authors acknowledge the support by Qualcomm Innovation Fellowship (QInF) 2012 and AFOSR MURI FA9550-09-1-0705.

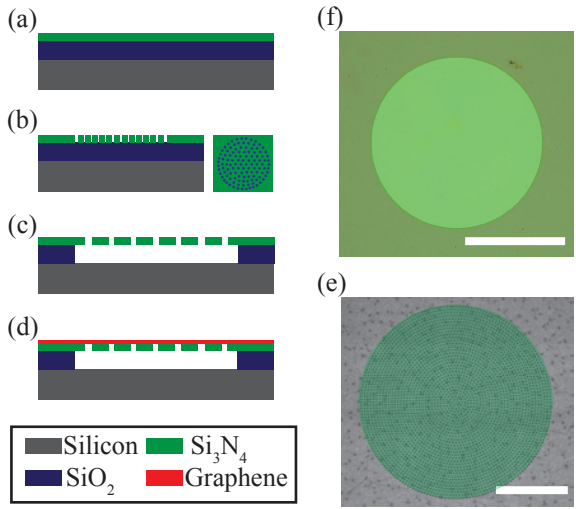

FIG 1. (a)-(d) Schematic of fabrication processes of $Si_3N_4$-graphene resonator: (a) $Si_3N_4$ deposition, (b) $Si_3N_4$ etch, (c) $SiO_2$ etch (d) graphene transfer. (e) optical image of graphene-coated resonator with 182 μm diameter. Scale bar: 100 μm. (f) false-color SEM image of a typical device. Scale bar: 50 μm.

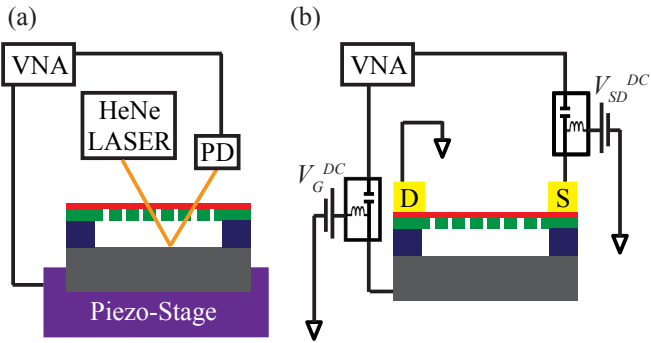

FIG 2. (a) Inertial drive and optical detection measurement scheme. (b) Electrical measurement scheme with shadow mask-defined source and drain electrodes.

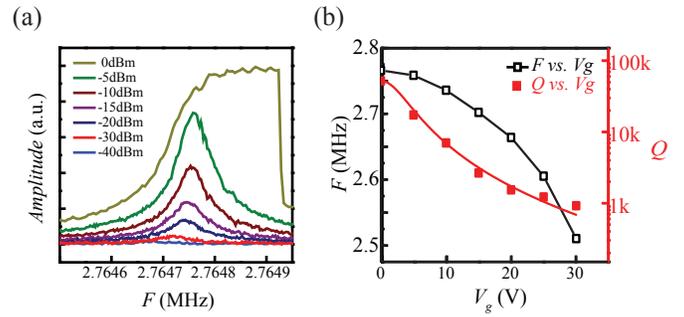

FIG 4. (a) Power handling of $Si_3N_4$ - graphene resonator. As the input power is increased, the resonance becomes nonlinear. (b) Gate tunability of $Si_3N_4$ - graphene resonator. The resonance down-shifts with increasing gate bias due to capacitive softening. The quality factor also degrades with gate bias due to the increase in displacement current as previously reported[15].

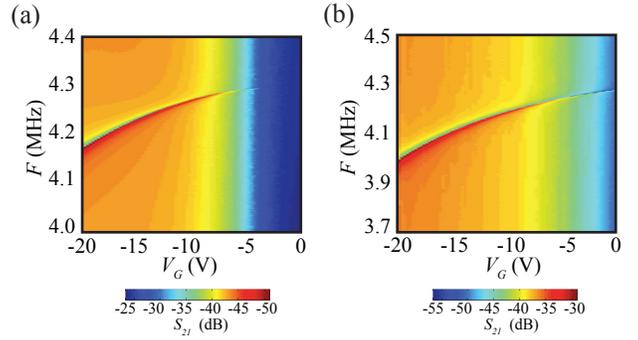

FIG 5. Fully-electrically integrated 91 μm SiNG drum resonator with 0 dBm source power. Both capacitive detection scheme (a) as well as direct detection scheme[14] with $V_{SD}$ = 9V (b) are implemented.

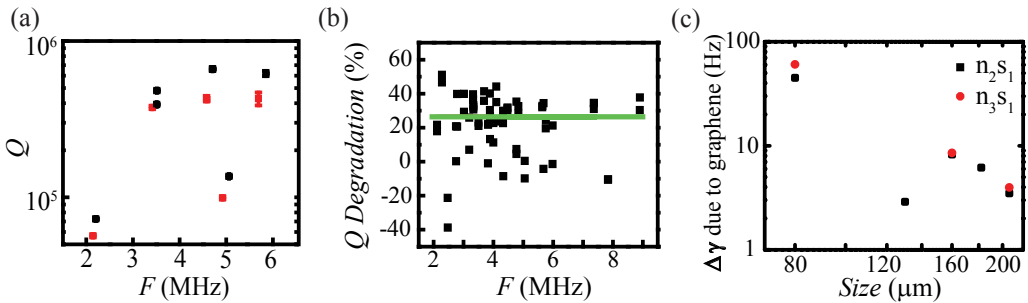

FIG 3. (a) Quality factor comparison for 206 μm $Si_3N_4$ drum resonator before (black) and after (red) graphene trasnfer. (b) Quality factor degradation measured on first five modes of 12 different SiNG drums with average (green line) ~26 %. (c) Change in damping rate for higher order modes due to graphene as a function of device dimensions.

# Supplementary materials for:

# Graphene Metallization of High-Stress Silicon Nitride Resonators for Electrical Integration

Sunwoo Lee[1, a], Vivekananda P. Adiga[2, a], Robert A. Barton[3, a], Arend van der Zande[3], Gwan-Hyoung Lee[4,5], B. Rob Ilic[6], Alexander Gondarenko[4], Jeevak M. Parpia[2], Harold G. Craighead[2], James Hone[4, b]

1. Department of Electrical Engineering, Columbia University, New York, New York 10027, USA.

2. School of Applied and Engineering Physics, Cornell University, Ithaca, New York 14853, USA.

3. Energy Frontier Research Center, Columbia University, New York, New York 10027, USA.

4. Department of Mechanical Engineering, Columbia University, New York, New York 10027, USA.

5. Samsung-SKKU Graphene Center (SSGC), Suwon, Gyeonggi 440-746, Korea.

6. The Cornell NanoScale Science & Technology Facility (CNF), Ithaca, New York 14853, Korea.

**Supplementary Materials**

The supplementary materials contain the following information:

1. Graphene growth and Raman measurements on CVD graphene on silicon nitride

2. Model fitting for Q vs. $V_{gate}$ analysis.

a) Sunwoo Lee, Vivekananda P. Adiga, and Robert A. Barton contributed equally to this work
b) Author to whom correspondence should be addressed.  Electronic mail: jh2228@columbia.edu

# 1. Graphene growth and Raman measurements on CVD graphene on silicon nitride

Small grain ( 1 ~ 5 µm) graphene is grown on copper foil[1] at 1000 °C with 35 sccm of methane flow and 2 sccm of hydrogen flow at 300 mTorr for 30 minutes after 30 minutes of hydrogen anneal (2 sccm) at 1000 °C. At the end of the growth cycle, the chamber is rapidly cooled to room temperature.

For large grain ( > 100 µm) graphene, a strip of copper if placed inside a crucible which is then sealed by a copper foil to reduce the flow into the copper strip[2]. The foil-sealed crucible containing the strip is then placed in a growth chamber where the growth is done at 1075 °C with 1 sccm of methane flow and 100 sccm of hydrogen flow at 1 Torr for 6 hours after 2 hours of hydrogrogen anneal (2 sccm) at 1075 °C. At the end of the growth cycle, the chamber is rapidly cooled to room temperature.

Figure S1 shows a Raman shift for the graphene transferred onto the silicon nitride substrate. While there is a uniform slope on the background from silicon nitride, no D peak and large 2D to G ratio is observed for both small grain and large grain graphene.

# 2. Model fitting for Q vs. Vgate analysis.

Due to the large resistance in graphene channel, displacement current induced dissipation is present which increases with increase gate bias, and such dissipation can be modeled as follows[3]

$$\frac{1}{Q(V_{dc})} = \frac{1}{Q_m} + \frac{\alpha V_{dc}^2}{1 + \beta |V_{dc}|} \qquad (S1)$$

$Q_m$ is the quality factor when there is no bias is applied, $V_{dc}$ is $V_{gate}$, and α parameterizes the displacement current from the change in gate-to-graphene capacitance while β parameterizes transconducting nature of graphene.

In detail, DC transport measurement can lead to finding β value as

$$R_g(V_{dc}) = [g_0(1 + \beta |V_{dc}|)]^{-1} \qquad (S2)$$

where $g_0$ is the conductance at $V_{gate} = 0$. Figure S2 shows a DC transfer of SiNG resonator with the field effect mobility of about 2000 cm$^2$/V-s, and corresponding fit using (S2) with $\frac{1}{g_0} = 4700\ \Omega$ and $\beta = -0.01$.

Finally, α is defined as

$$\alpha = \left(\frac{dC}{dD}\right)^2 \bigg|_{D=D_{eq}} / m_{eff} \omega_m g_0 \tag{S3}$$

where D is the distance between the gate and the graphene. Since a SiNG drum contains two different dielectric materials, silicon nitride and air, one can write

$$\frac{d}{dD}\left(\frac{1}{C}\right) = \frac{d}{dD}\left(\frac{1}{C_{air}} + \frac{1}{C_{Si_3N_4}}\right) \tag{S4}$$

$$\frac{1}{C^2}\frac{dC}{dD} = \frac{1}{C_{air}^2}\frac{dC_{air}}{dD} + \frac{1}{C_{Si_3N_4}^2}\frac{dC_{Si_3N_4}}{dD} \tag{S5}$$

Since $C_{Si_3N_4}$ does not vary during resonance and air-capacitance is dominant, (S4) simplifies

$$\frac{dC}{dD} = \frac{C^2}{C_{air}^2}\frac{dC_{air}}{dD} \sim \frac{dC_{air}}{dD} \tag{S6}$$

In addition, as r (62 μm) >> d (670 nm), the circular plate capacitance can be approximated as a parallel plate capacitor. It should be noted that d is about 670 nm because of $Si_3N_4$ has finite etch rate against HF, and hence is thinned down during the HF release step.

$$C_{plate} = \frac{\varepsilon_0 \varepsilon_r A}{d} \tag{S7}$$

With the parameters obtained from the above and $m_{eff}$ of 4 times the mass calculated from the density and dimension of the silicon nitride are used to fit the data shown in the Fig 4(b).

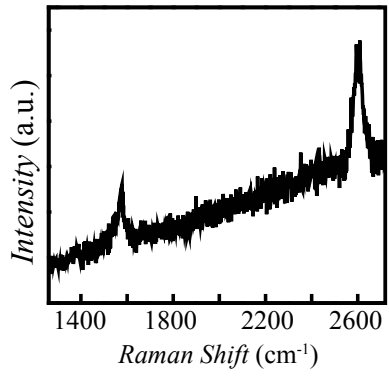

FIG S1. (a) Raman spectrum of suspended graphene - $Si_3N_4$ resonator.

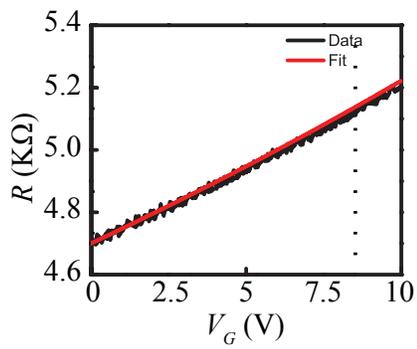

FIG S2. DC transport a typical SiNG resonator ($\mu_{FE} \sim 2000$ cm$^2$/V-s) and fit to find $\beta$.